\documentclass[10pt,conference]{IEEEtran}
\usepackage{amsmath}
\usepackage{cite}
\usepackage{graphicx}
\usepackage{amsfonts}
\usepackage{latexsym}
\usepackage{subfigure}
\usepackage{algorithm}
\usepackage{algorithmic}
\usepackage{color}
\usepackage{times}
\usepackage{epsfig, graphics}
\usepackage{bm}
\usepackage{subfigure}
\usepackage{graphicx,wrapfig,lipsum}

\begin{document}

\newtheorem{theorem}{Theorem}
\newtheorem{lemma}{Lemma}
\newtheorem{conjecture}{Conjecture}
\newtheorem{corollary}{Corollary}
\newtheorem{definition}{Definition}
\newtheorem{scheme}{Scheme}
\newcommand{\argmax}{\arg\!\max}
\newcommand{\pound}{\operatornamewithlimits{\gtrless}}
\IEEEoverridecommandlockouts

\title{Spectrum Data Poisoning with Adversarial Deep Learning
\thanks{This effort is supported by the U.S. Army Research Office under contract
W911NF-17-C-0090. The content of the information does not necessarily
reflect the position or the policy of the U.S. Government, and no official
endorsement should be inferred.}}
\author{Yi Shi, Tugba Erpek, Yalin E. Sagduyu, and Jason H. Li\\
Intelligent Automation, Inc., Rockville, MD, USA\\
\{yshi, terpek, ysagduyu, jli\}@i-a-i.com
}

\maketitle

\begin{abstract}
Machine learning has been widely applied in wireless communications. However, the security aspects of machine learning in wireless applications have not been well understood yet. We consider the case that a cognitive transmitter senses the spectrum and transmits on idle channels determined by a machine learning algorithm. We present an adversarial machine learning approach to launch a spectrum data poisoning attack by inferring the transmitter's behavior and attempting to falsify the spectrum sensing data over the air. For that purpose, the adversary transmits for a short period of time when the channel is idle to manipulate the input for the decision mechanism of the transmitter. The cognitive engine at the transmitter is a deep neural network model that predicts idle channels with minimum sensing error for data transmissions. The transmitter collects spectrum sensing data and uses it as the input to its machine learning algorithm. In the meantime, the adversary builds a cognitive engine using another deep neural network model to predict when the transmitter will have a successful transmission based on its spectrum sensing data. The adversary then performs the over-the-air spectrum data poisoning attack, which aims to change the channel occupancy status from idle to busy when the transmitter is sensing, so that the transmitter is fooled into making incorrect transmit decisions. This attack is more energy efficient and harder to detect compared to jamming of data transmissions. We show that this attack is very effective and reduces the throughput of the transmitter substantially.

\end{abstract}


\section{Introduction}
Cognitive radios support the efficient discovery and use of the spectrum \cite{Mitola}. Spectrum sensing is one of the key tasks of cognitive radios to achieve situational awareness. Cognitive radios analyze the signals received through spectrum sensing and determine their channel access strategy accordingly. Because of the open and broadcast nature of wireless communications, the decisions built upon spectrum sensing results are susceptible to different types of attacks that aim to force the cognitive radios to make incorrect decisions.

Different attacks on the spectrum sensing decisions in a cognitive network have been studied in the literature and various defense techniques have been developed \cite{Clancy08:CogSec}.
\begin{itemize}
\item \emph{Primary user emulation} (PUE) attacks decrease the spectrum access opportunities of cognitive radios.  A defense mechanism with belief propagation was studied in \cite{PUE}.
\item  In a collaborative sensing environment, some cognitive users may send falsified reports to a decision center. This corresponds to a \emph{spectrum sensing data falsification} (SSDF) attack  that aims to degrade the performance of spectrum sensing. A trust-based defense strategy to mitigate this attack was studied in \cite{Sagduyu2014} by using a common receiver for data fusion. SSDF attack also applies to mobile ad hoc networks (MANETs) where there is no centralized data fusion center. A consensus-based cooperative spectrum sensing scheme to counter SSDF attacks in cognitive radio MANETs was studied in \cite{Yut}.
\item Cognitive radio networks are also susceptible to conventional security threats such as \emph{jamming} \cite{Sagduyu11:jamming}, cognitive interference \cite{Sagduyu118:satellite}, \emph{eavesdropping} \cite{Zou15:eavesdropping} and \emph{noncooperation} \cite{Sagduyu09:noncoop}. These threats extend from physical layer to higher layers (e.g., routing in the network layer \cite{Lu2014, Lu15:Attack}), and exploit different levels of uncertainty regarding channel, traffic, and adversary types \cite{Attack1, Sagduyu10:Attack, Sagduyu07:Attack, Attack5}.
\end{itemize}

In this paper, we introduce a new type of attack motivated by adversarial machine learning, namely the \emph{over-the-air spectrum data poisoning attack}. While the adversary jams the channel under this attack, its purpose is not to degrade the data transmission received (as typically assumed in denial-of-service attacks \cite{DOS}) but it aims to manipulate the spectrum sensing data collected so that wrong transmit decisions are made by using the unreliable spectrum sensing results. Also, this attack differs from the SSDF attack, since the adversary does not participate in cooperative spectrum sensing and does not try to change channel labels directly as in the SSDF attack. Instead, the adversary injects adversarial perturbation to the channel in order to fool the transmitter into making wrong transmit decisions.

Recently, machine learning has been applied for several cognitive radio tasks such as spectrum sensing \cite{Kemal2018}, spectrum access \cite{Yi2018, Tugba2018} and  modulation recognition \cite{OShea2016}.
However, there are various security concerns regarding the safe use of machine learning algorithms. For example, if the input data to a machine learning algorithm is manipulated during the training or operation (test) time, the output will be very different compared to the expected results. \emph{Adversarial machine learning} studies learning in the presence of adversaries and aims to enable safe adoption of machine learning to the emerging applications.

Three broad categories of attacks under adversarial machine learning are  \emph{exploratory (or inference) attacks}, \emph{evasion attacks} and \emph{causative  (or poisoning) attacks}.
\begin{itemize}
\item In \emph{exploratory (or inference) attacks} \cite{Ateniese,Tramer,Fredrikson,Shi17:HST}, the adversary aims to understand how the underlying machine learning works for an application (e.g., inferring sensitive and/or proprietary information).
\item In \emph{evasion attacks} \cite{Biggio,Kurakin}, the adversary attempts to fool the machine learning algorithm into making a wrong decision (e.g., fooling a security algorithm into accepting an adversary as legitimate).
\item In \emph{poisoning  (or causative) attacks} \cite{Biggio2, Pi}, the adversary provides incorrect information such as training data to machine learning.
\end{itemize}
These attacks can be launched separately or combined, i.e., causative and evasion attacks can be launched by making use of the inference results from an exploratory attack \cite{ShiMilcom,Yibook}.

In this paper, we apply \emph{adversarial deep learning} to launch an \emph{exploratory attack} on a cognitive radio as the preliminary step before intentionally changing the transmitter's sensing results by transmitting when there will be a success transmission if this transmission is not under attack.
For that purpose, the adversary trains a deep neural network.
We consider a canonical wireless communication scenario with a cognitive transmitter, the corresponding receiver, an adversary, and some other background traffic.
The cognitive transmitter builds a machine learning model (based on a deep neural network) to predict the busy and idle states of the channel. The training data includes
\begin{itemize}
	\item time-series of spectrum sensing results as \emph{features}, and
	\item channel idle/busy status based on the ground truth (the background transmitter's on/off state) as \emph{labels}.
\end{itemize}
Then this machine learning model is used by the cognitive transmitter to make transmit decisions.
If a transmission is successful (i.e., the signal-to-interference-plus-noise ratio (SINR) exceeds a threshold), the receiver sends an acknowledgement (ACK) to the transmitter, which can also be overheard by an adversary.
The adversary performs an \emph{exploratory attack} to build a classifier that can predict the outcome of transmissions, i.e., whether there will be an ACK or not if no attack.
Note that this is not a standard exploratory attack and the classifier built by the adversary will not be the same as (or similar to) the classifier used by the transmitter, due to the following two differences.
\begin{itemize}
	\item The transmitter and the adversary are in different locations and thus their sensing results will vary based on the channel environment and differ from each other. As a result, the input data to their classifiers will differ.
	\item The adversary predicts the outcome of transmissions (`ACK' or `no ACK') while the transmitter predicts channel status (idle or busy). As a result, the output data of their classifiers will differ.
\end{itemize}

After building its classifier, the adversary predicts when the transmitter will have a successful transmission (if no attack) and performs the \emph{poisoning attack}, i.e., the adversary transmits to change the channel status in order to poison (i.e., falsify) the transmitter's input (spectrum sensing data) to the machine learning algorithm.
The attack considered in this paper is similar to that in \cite{Yi2018}, where the adversary also first learns the transmitter's behavior (ACK or not) by an exploratory attack and then performs subsequent attacks.
The difference is that in \cite{Yi2018}, the adversary performs a jamming attack during the data transmission phase to make a transmission fail while in this paper the adversary performs a spectrum data poisoning attack in the sensing phase such that the transmitter has incorrect input data to its classifier and makes the wrong decision of not transmitting.
The attack considered in this paper is hard to detect since it does not directly jam the transmitter's signal but it changes the input data to the decision mechanism so that the transmitter chooses not to transmit when the channel is indeed idle.
Moreover, this attack is energy efficient since the adversary makes a very short transmission in the sensing period.

We show that this adversarial deep learning approach results in an effective attack. In particular, for the scenario studied in numerical results, only few transmission attempts are made and the achieved throughput (normalized by the optimistic throughput by an ideal algorithm to detect every idle channel) drops from $98.96\%$ to $3.13\%$, when the spectrum data poisoning attack is launched.

The rest of the paper is organized as follows.
Section~\ref{sec:scenario} describes the system model.
Section~\ref{sec:transmitter} describes the transmitter's algorithm and shows the performance without an attack.
Section~\ref{sec:adversary} describes the adversary's algorithm and shows the performance under the attack.
Section~\ref{sec:conclusion} concludes the paper.

\section{System Model}
\label{sec:scenario}

\begin{figure}
	\centering
	\includegraphics[width=0.9\columnwidth]{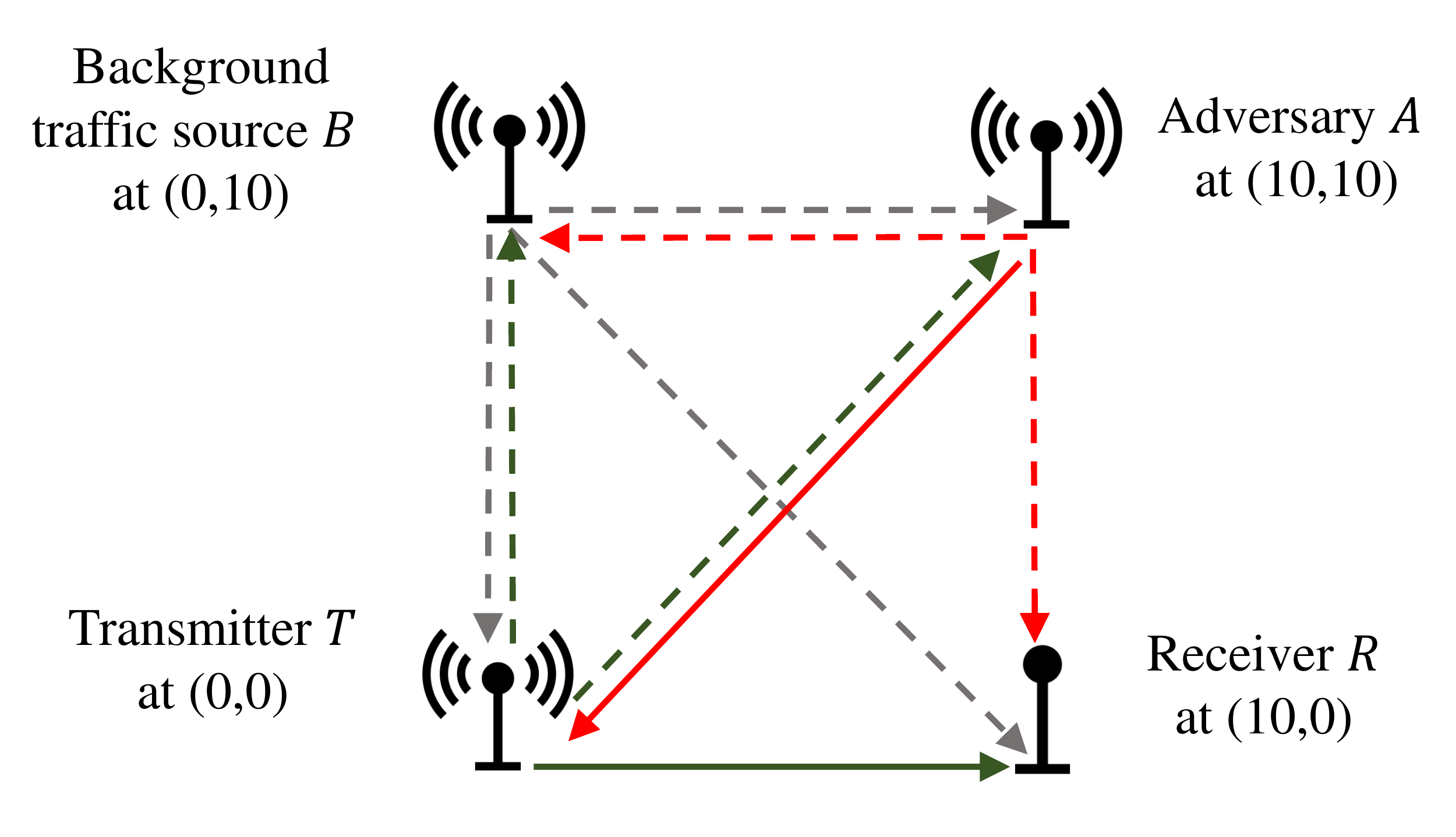}
	\caption{The network topology.}
	\label{fig:topology}
\end{figure}

We consider a cognitive network that includes a transmitter $T$, a receiver $R$, an adversary $A$ and some background traffic source $B$  that may transmit its data. The network topology to generate numerical results is shown in Figure~\ref{fig:topology}.
Note that the designed attack schemes can be directly applied to other network topologies.

The activities of $B$ are not known \emph{a priori} and can be detected via spectrum sensing. Time is divided in slots. Within each slot, the initial short period of time is allocated by $T$ for spectrum sensing and the ending short period of time is allocated for feedback (ACK).
The rest of a slot is for data transmission if no other transmission is detected.
The decision of  $T$ is based on a deep neural network (trained by deep learning) that analyzes sensing results and then determines the channel status such that the channel is busy if background traffic is detected and channel is idle otherwise.
Each sensing result is either
\begin{itemize}
	\item noise with normalized power (when channel is idle) or
	\item noise plus the received power from background traffic (when channel is busy).
\end{itemize}	
	We assume that the noise and the received power are random variables with Gaussian distributions.

Data transmission is successful if the SINR at the receiver $R$ is not less than some threshold, where the interference comes from the transmissions of $B$.
We assume Gaussian noise at $R$ and Gaussian channel gain from $T$ to $R$. The mean value of the channel gain is calculated based on the free-space propagation loss model.
A receiver sends an ACK for each successful transmission.
The adversary $A$ also senses the channel and aims to predict whether there will be a successful transmission (ACK) if no attack.
Note that $A$ only detects the presence of the ACK message but does not need to decode it.
The prediction by $A$ is based on another deep neural network that is trained by $A$ using deep learning.
If $A$ predicts that there will be a successful transmission, it performs an attack to reduce the throughput of $T$.

In this paper, we consider the attack of transmitting in the initial short sensing period to change the sensing result of $T$ for the current time slot.
Since this sensing result is an input to $T$'s classifier on channel status, $T$ may make a wrong decision, even if its classifier algorithm was built well.

The advantage of this attack, comparing with the continuous jamming attack, is that the initial sensing period is much shorter than the data transmission period.
As a result, the power consumption of this attack is also much less compared to continuous jamming. In addition, this attack is harder to detect compared to continuous jamming.

We can design a defense mechanism to prevent such attacks by noting that the first step of the attack is an exploratory attack to understand  $T$'s classifier.
Thus,  $T$ can take wrong actions in a controlled manner such that the `ACK' or `no ACK' results can be changed.
As a consequence, the classifier built by an adversary may not work well and thus the attack may become less effective.

\section{Transmitter's Algorithm}

\label{sec:transmitter}
Transmitter $T$ needs to sense the spectrum, identify an idle channel (the other transmitter is not transmitting), and then decide when to transmit.  $T$ applies deep learning to train a deep neural network classifier that identifies idle channels.
This classifier is pre-trained using the most recent $10$ sensing results as features and the current channel busy/idle status as a label to build one training sample. The number of sensing results used for one sample is potentially a design parameter for $T$ and can be tuned by $T$ to optimize its performance.
Each sensing result is either a Gaussian noise with normalized power $1$ (when channel is idle) or noise plus the transmit power from another user $g_{BT} P$ (when channel is busy), where $g_{BT}$ is the channel gain from the other transmitter $B$ to $T$ and $P$ is the transmission power (noise and the channel gain are random variables with Gaussian distributions).
After observing certain period of time, $T$ collects a number of samples to be used as training to build a deep learning classifier.

\begin{figure}
	\centering
	\includegraphics[width=0.55\columnwidth]{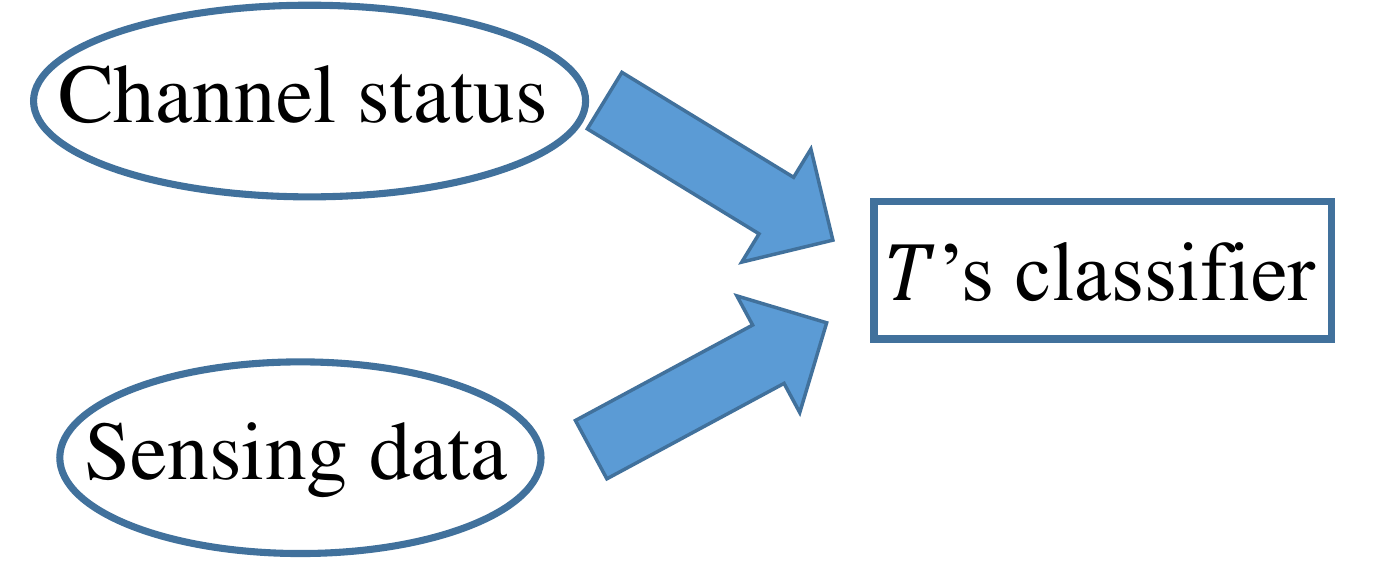}
	\caption{The input and output (label) data while training $T$'s classifier.}
	\label{fig:tsclassifiertraining}
\end{figure}

Figure~\ref{fig:tsclassifiertraining} shows the input data and the labels while building $T$'s classifier. $T$'s training algorithm is summarized as follows.
\begin{itemize}
\item $T$ collects data over a time period. Denote the sensed power at time $t$ as $p_t$.

\item $T$ builds a training sample $\{(p_{t-9}, p_{t-8}, \cdots, p_t), s_t\}$ for each time $t \ge 10$, where $s_t$ is the channel busy/idle condition at time $t$.

\item $T$ divides samples equally to a training set and a test set and uses one half to train a classifier with deep learning.
\end{itemize}

Once a classifier is built, $T$ will use it to predict the channel status and transmit if it predicts the channel as idle.
This prediction algorithm is summarized as follows.
\begin{itemize}
\item At time $t$, $T$ senses channel and obtains power $p_t$. $T$ builds a test sample $(p_{t-9}, p_{t-8}, \cdots, p_t)$.

\item $T$ uses its classifier to decide on the labels (busy or idle) and then transmits if the channel is predicted idle.
\end{itemize}
For this algorithm, there may be two types of errors:
\begin{itemize}
\item \emph{Misdetection}. The channel is busy but it is detected as idle.

\item \emph{False alarm}. The channel is idle but it is detected as busy.
\end{itemize}
Transmitter $T$ aims to minimize $\max\{ e_{MD}, e_{FA} \}$, where $e_{MD}$ and $e_{FA}$ are the probabilities of misdetection and false alarm at $T$, respectively.
These error probabilities are calculated by $e_{MD} = \frac{n_{MD}}{n_{busy}}$ and $e_{FA} = \frac{n_{FA}}{n_{idle}}$, where $n_{MD}$ is the number of misdetections, $n_{busy}$ is the number of times that the channel is busy, $n_{FA}$ is the number of false alarms, and $n_{idle}$ is the number of times that the channel is idle.

\begin{figure}
	\centering
	\includegraphics[width=\columnwidth]{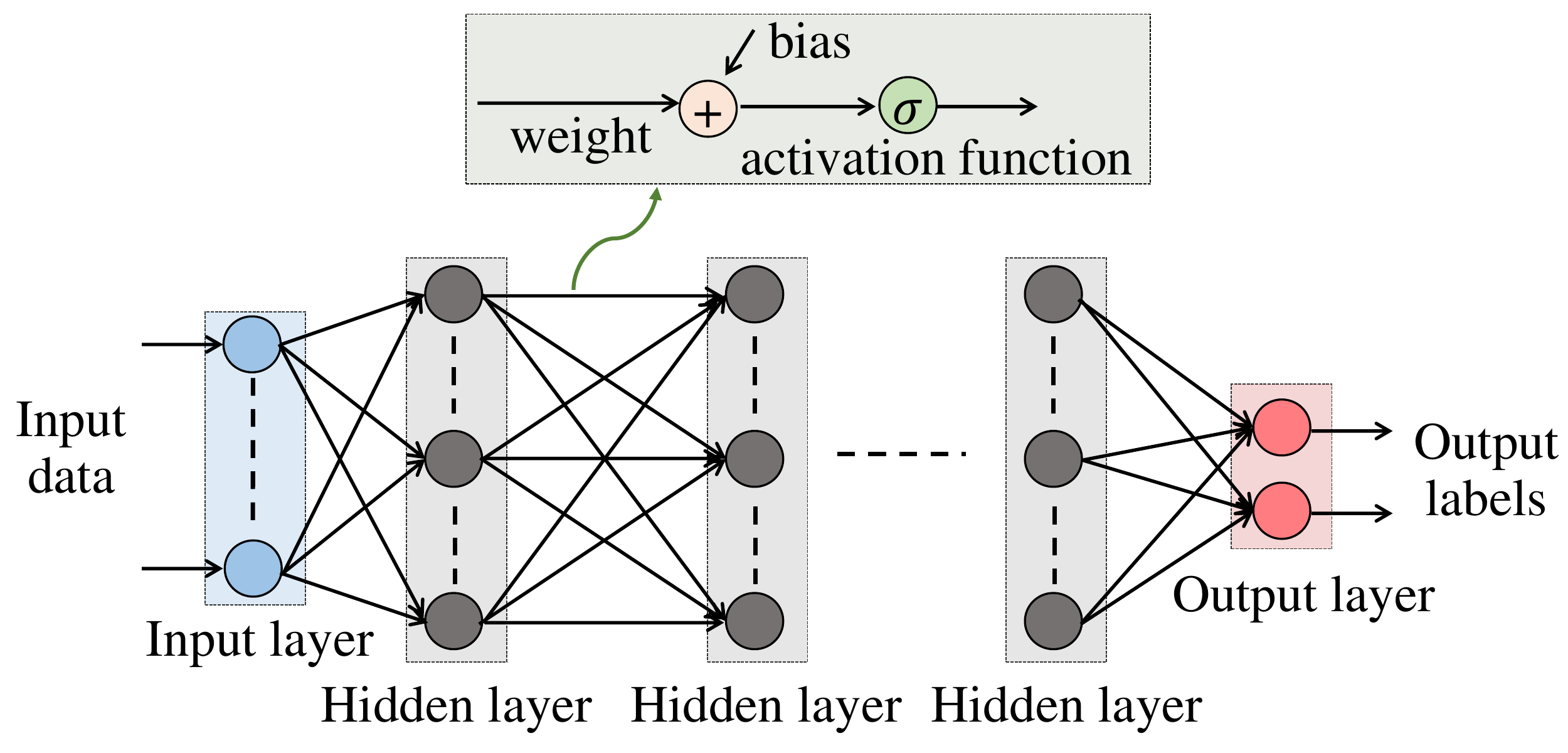}
	\caption{The structure of a feedforward neural network.}
	\label{fig:fnn}
\end{figure}

We use TensorFlow to build a deep learning classifier for $T$.
In particular, we use the following deep neural network:
\begin{itemize}
	\item A feedforward neural network is trained with backpropagation function by using cross-entropy as the loss function. The structure of a feedforward neural network is shown in Figure~\ref{fig:fnn}.
	\item Number of hidden layers is 3.
	\item Number of neurons per hidden layer is 100.
	\item Rectified linear unit (ReLU) is used as activation function at hidden layers.
	\item Softmax is used as the activation function at output layer.
	\item Batch size is 100.
	\item Number of training steps is 1000.
\end{itemize}

\begin{figure}
	\centering
	\includegraphics[width=\columnwidth]{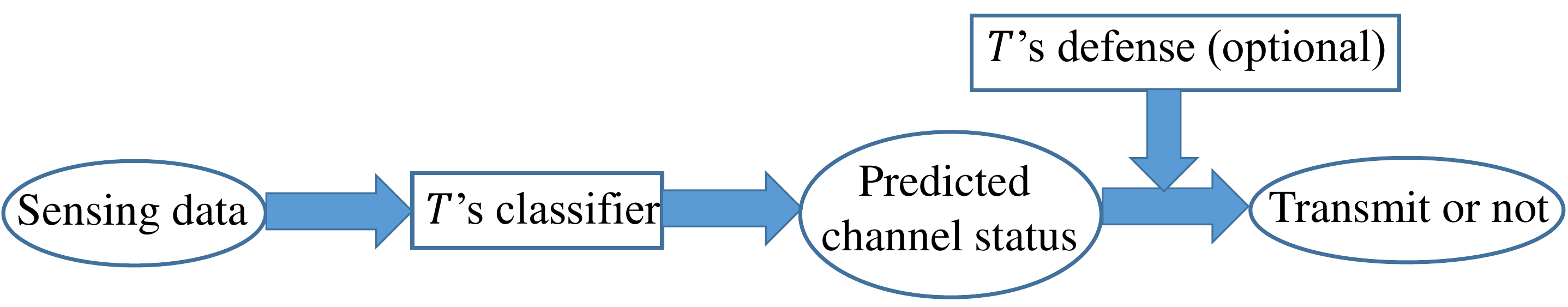}
	\caption{$T$'s classifier during run-time.}
	\label{fig:tsclassifierruntime}
\end{figure}

Note that $T$ can further optimize the hyperparameters (e.g., number of layers and number of neurons per layer) of its deep neural network. The block diagram in Figure~\ref{fig:tsclassifierruntime} shows $T$'s operation during the run-time.
Note that there is an optional block of defense, 
which was discussed in Section~\ref{sec:scenario}.

Background traffic arrives at another transmitter $B$ at rate of $0.8$ packet per time slot.
When $B$ has queued data packet, it may decide to transmit at rate of $1$ packet per time slot and once it transmits, it will continue until the queue is empty.
The channel gain $g_{BT}$ is a random variable with a Gaussian distribution and the expected value $d_{BT}^{-2}$, where $d_{BT}$ is the distance between $B$ and $T$.
In the simulation setting, the location of $B$ is $(0,10)$, the location of $T$ is $(0,0)$, and the transmit power at $B$ is $1000$ (normalized with respect to the unit noise power).

$T$ collects $1000$ samples (each with the most recent $10$ spectrum sensing results and a label (`idle' or `busy') and uses half of them as training data and the other half as test data.
There are $403$ busy and $97$ idle channels in the test data.
Among them, $3$ busy channels are identified as idle.
Thus,
\begin{eqnarray*}
	&& e_{FA}  = 0\%, \\ && e_{MD} = 3/403 = 0.74\%.
\end{eqnarray*}
Both errors are small, showing that $T$ can reliably predict the channel status.

$T$ transmits data in idle channels detected by its deep learning classifier.
If the SNR (or SINR) at receiver $R$ is no less than some threshold, $R$ will confirm a successful transmission by sending an ACK to $T$.
Note that we again assume a Gaussian noise at $R$ and a Gaussian channel gain from $T$ (or $B$) to $R$.

In the simulation, we set threshold as $3$, the location of $R$ as $(10,0)$, and the transmit power at $T$ as $1000$ (normalized with respect to the unit noise power).
 $T$ applies its deep learning classifier on $500$ time slots and makes transmission decisions.
There are $404$ busy and $96$ idle channel instances in these time slots.
Among them, $2$ busy channel instances are identified as idle and transmissions in these $2$ slots fail, while $96$ idle channel instances are correctly identified as idle and transmissions in $95$ slots are successful.

To measure the throughput performance of $T$'s algorithm, we compare it with an ideal algorithm that can correctly detect all idle channels.
Thus, the achieved normalized throughput $t$ is defined as the ratio of the number of successful transmissions to the number of slot with idle channel instances. Without an attack, $t$ is measured as
\begin{eqnarray*}
t = 95/96=98.96\%.
\end{eqnarray*}
We also evaluate the success ratio $r$, which is defined as the ratio of the number of successful transmissions to the number of all transmissions. Without an attack, $r$ is measured as
\begin{eqnarray*}
s = 95/(96+2)=96.94\%.
\end{eqnarray*}
We can see that due to small errors to detect busy/idle channels, $T$'s algorithm achieves high values of normalized throughput and success ratio. Finally, we evaluate the all transmission ratio $a$, which is defined as the ratio of the number of all transmissions to the number of all slots. Without an attack, $a$ is measured as
\begin{eqnarray*}
	a = (96+2)/500 = 19.60\%.
\end{eqnarray*}

\section{Adversary's Algorithm}
\label{sec:adversary}

There is an adversary $A$ that aims to reduce transmitter $T$'s performance.
As the first step, it needs to launch an exploratory attack to infer $T$'s classifier.
In particular, $A$ senses the spectrum, predicts whether there will be a successful transmission (if no attack), and performs certain attacks if there will be a successful transmission.
There are four cases:
\begin{enumerate}
	\item channel is idle and $T$ is transmitting,
	\item channel is busy and $T$ is not transmitting,
	\item channel is idle and $T$ is not transmitting, or
	\item channel is busy and $T$ is transmitting.
\end{enumerate}
Ideally, the last two cases should be rare cases. We assume that $A$ can hear ACKs for $T$'s transmissions (but does not decode them). $A$ uses the most recent $10$ sensing results as the features and the current feedback (`ACK' vs. `no ACK') as the label to build one training sample. The number of sensing results used for one sample is potentially a design parameter for $A$ and can be tuned by $A$ to optimize the impact of its attack.

\begin{figure}
	\centering
	\includegraphics[width=\columnwidth]{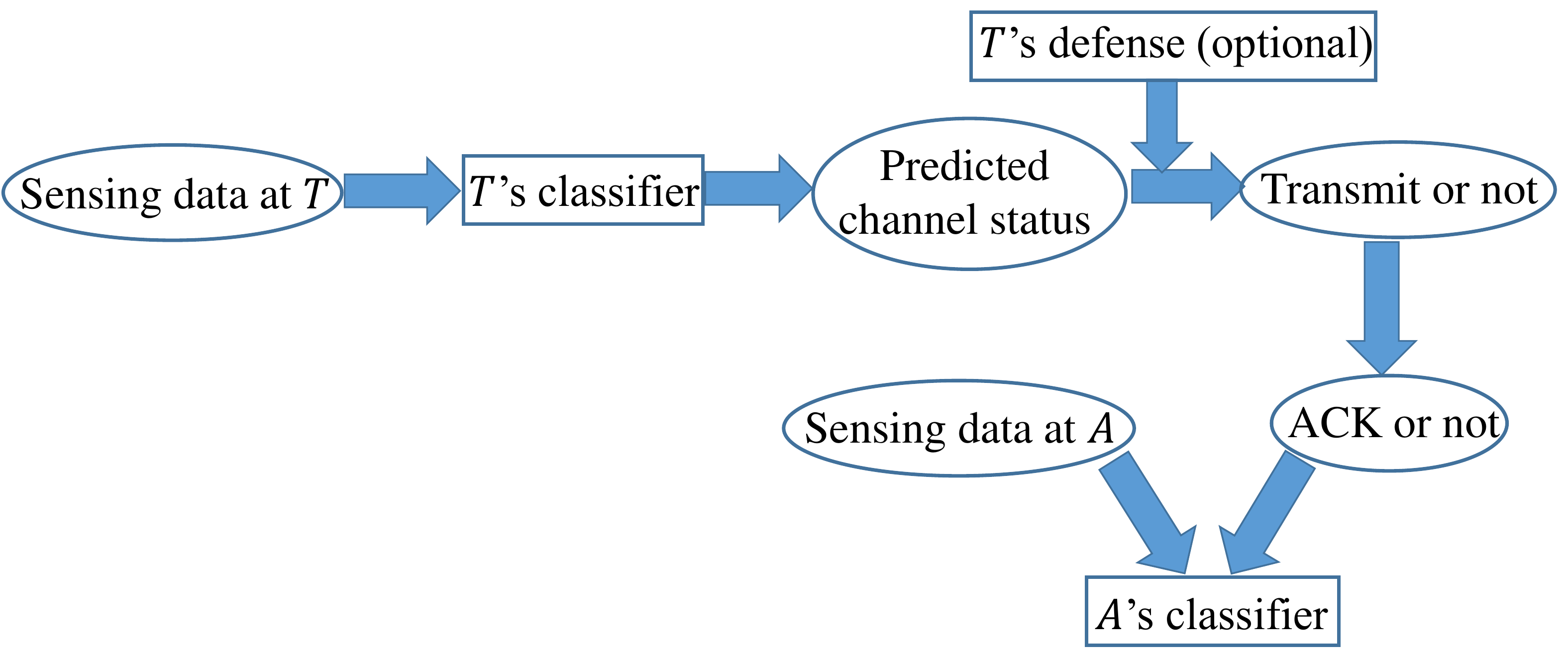}
	\caption{The input and output (label) data while training $A$'s classifier.}
	\label{fig:asclassifiertraining}
\end{figure}

After observing certain period of time, $A$ collects a number of samples as training data to build a deep learning classifier that outputs one of two labels,  `a successful transmission' and `failed transmission'.
Figure~\ref{fig:asclassifiertraining} shows the input data and the labels while building the adversary's classifier. $A$ uses the same deep neural network structure as $T$, although it determines its own weights and biases of the deep neural network by using its own training data. $A$ can further optimize the hyperparameters of its deep neural network.
$A$'s training algorithm is summarized as follows.
\begin{itemize}
\item $A$ collects data over a time period. Denote the sensed power at time $t$ as $p_t$ and the label (ACK or not) at time $t$ as $l_t$.

\item $A$ builds a training sample $\{(p_{t-9}, p_{t-8}, \cdots, p_t), l_t\}$ for each time $t \ge 10$.

\item $A$ divides samples equally to a training set and a test set and one half to train a classifier using deep learning.
\end{itemize}

The process of building such a classifier can be regarded as an \emph{exploratory attack}, since $A$ aims to build a classifier to infer the operation of $T$. The classifier built by $A$ is similar to $T$'s classifier. However, there are the following two differences between the classifiers built by $T$ and $A$.
\begin{itemize}
\item Due to different locations of $T$ and $A$ and random channels, the sensing results at $T$ and $A$ differ.
Thus, features for the same sample are different at $T$ and $A$.

\item The labels (classes) are different, i.e., `busy' or `idle' in $T$'s classifier and `ACK' or `no ACK' in $A$'s classifier.
\end{itemize}

Once a classifier is built, $A$ uses it to predict whether there is a successful transmission (if no attack).
This prediction algorithm is given as follows.
\begin{itemize}
\item At time $t$, $A$ senses channel and obtains power $p_t$. $A$ builds a test sample $(p_{t-9}, p_{t-8}, \cdots, p_t)$.

\item $A$ uses its classifier to decide on a label (ACK or not) and then attacks if there will be an ACK.
\end{itemize}
For this algorithm, there may be two types of errors:
\begin{itemize}
\item \emph{Misdetection}. There will be a successful transmission but $A$'s classifier predicts that there will not be a successful transmission.

\item \emph{False alarm}. There will not be a successful transmission but $A$'s classifier predicts that there will be a successful transmission.
\end{itemize}
$A$ aims to minimize $\max\{\tilde{e}_{MD}, \tilde{e}_{FA} \}$, where $\tilde{e}_{MD}$ and $\tilde{e}_{FA}$ are the probabilities of misdetection and false alarm at $A$, respectively.
We use TensorFlow to build a deep learning classifier for $A$. The deep neural network structure is the same as the one used by $T$ (described in Section~\ref{sec:transmitter}). As the features and labels are different for $T$ and $T$, they individually train their deep neural network (i.e., determine their own weights and biases).

\begin{figure}
	\centering
	\includegraphics[width=\columnwidth]{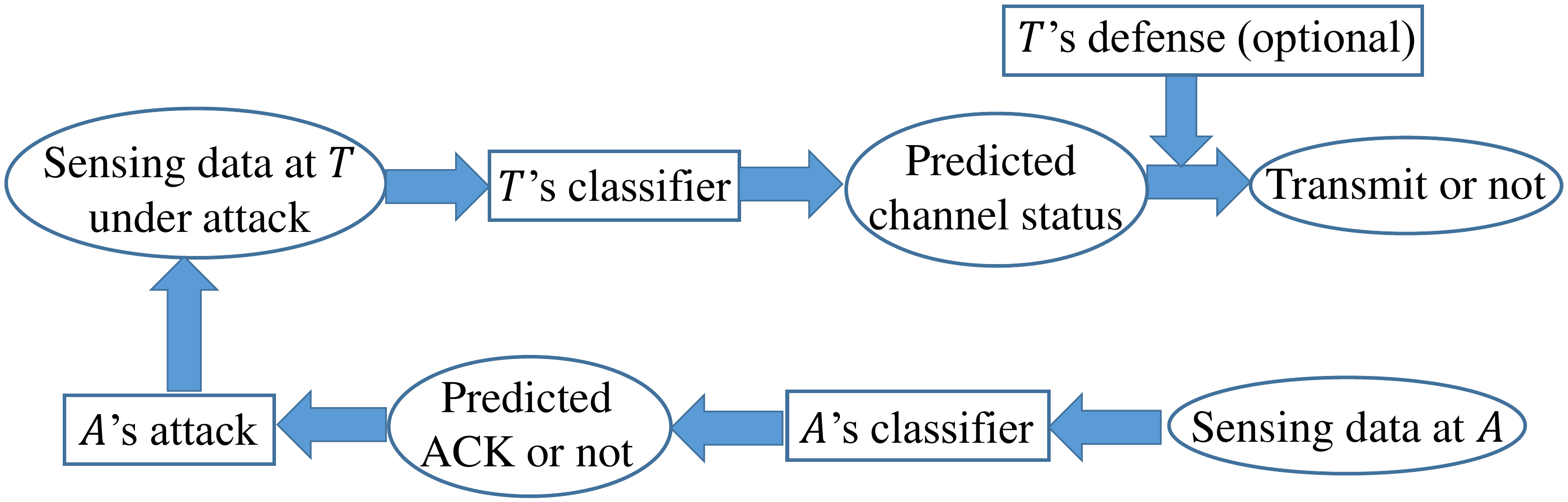}
	\caption{Using $A$'s classifier for attacks.}
	\label{fig:asclassifierruntime}
\end{figure}

Figure~\ref{fig:asclassifierruntime} illustrates the attacker's operation during the run-time.
In the simulation, we set the location of $A$ as $(10,10)$. $A$ collects $1000$ samples with labels and uses half of them for training data and the other half for test data.
There will be $95$ successful transmissions in $500$ test data. Out of $95$ transmissions, $4$ are predicted as failed transmissions, although these transmissions are indeed successful.
Among $405$ failed transmissions, $8$ of them are predicted as successful transmissions, although these transmissions indeed fail.
Thus,
\begin{eqnarray*}
&& \tilde{e}_{FA}  =  8/405=1.98\%, \\ && \tilde{e}_{MD}  =  4/95=4.21\%.
\end{eqnarray*}
Both errors are small, showing that $A$ can reliably predict the successful transmissions by $T$.

With the output of $A$'s classifier, $A$ performs an attack by transmitting in the initial short sensing period to change $T$'s sensing result for the current time slot.
This sensing result is one feature of $T$'s classifier and thus $T$ may make a wrong decision, even if its classifier algorithm was built well. Compared with a continuous jamming attack, this attack targets the initial sensing period that is much shorter than the data transmission period.
Hence, the power consumption of this attack is much less than continuous jamming.
In the simulation, the transmit power at $A$ is set as $1000$.

For the classifier built in Section~\ref{sec:transmitter} and $500$ time slots considered for transmissions under the attack, $1$ busy channel is identified as idle and the transmission in this slot fails, while $3$ (of $96$) idle channels are identified as idle and the transmissions in these $3$ slots are all successful.
Thus, the achieved normalized throughput is
\begin{eqnarray*}
t = 3/96=3.13\%,
\end{eqnarray*}
 and the overall success ratio is measured as
\begin{eqnarray*}
s = 3/(3+1)=75\%,
\end{eqnarray*}
while only very few transmission attempts are made such that the all transmission ratio $a$ is measured as
\begin{eqnarray*}
	a = (3+1)/500=0.80\%.
\end{eqnarray*}

As a result, $A$ reduces the throughput of $T$ significantly from $98.96\%$ to $3.13\%$ and reduces
the success ratio from $96.94\%$ to $75\%$.
Table~\ref{table:attack} summarizes the performance of $T$ with and without an attack, and demonstrates the success of this attack.

\begin{table}
	\caption{Results with and without attack.}
	\centering
	{\small
		\begin{tabular}{c|c|c|c}
			& Normalized & Success & All transmission \\
			& throughput $t$ & ratio $s$ & ratio $a$ \\ \hline \hline
			no attack & 98.96\% & 96.94\% & 19.60\% \\ \hline
			with attack & 3.13\% & 75.00\% & 0.80\%
		\end{tabular}
	}
	\label{table:attack}
\end{table}

\section{Conclusion}
\label{sec:conclusion}

We applied adversarial machine learning (based on deep neural networks) to design an over-the-air spectrum sensing data poisoning attack that manipulates the input data of the transmitter during the run-time and fools it into making wrong transmit decisions.
The proposed attack is preferable for the adversary with a power constraint since the adversary only needs to transmit for a short period of time to manipulate the transmit decisions of the transmitter.
Moreover, it is not easy to detect such an attack due to the short period of transmit time.
We showed that this attack substantially decreases the throughput of the transmitter, while forcing it to make few transmission attempts. The results show the effectiveness of the proposed spectrum poisoning attack, which raises the need of new defense mechanisms to protect wireless communications against intelligent attacks based on adversarial machine learning.

\end{document}